# Alleviating Vulnerabilities of the Possible Outbreaks of Measles: A Data Trend Analysis and Prediction of Possible Cases

Hidear Talirongan[1], Markdy Y. Orong[2] & Florence Jean B. Talirongan[3]

[1,2,3]Misamis University, Ozamiz City, Philippines.

## ABSTRACT

Measles is considered as a highly contagious disease that leads to serious complications around the world. Thus, the paper determined the trend and the five-year forecasted data of the Measles in the Philippines. This study utilized the time series data for trend analysis and data forecasting using the ARIMA model to visualize the measles cases. Figures for the time-series and forecasted results are individually presented with the use of GRETL software. Results showed that there was an increasing pattern of the disease from 2016 to 2019. However, there was a decreasing pattern of its occurrence in the next five years based on the five-year forecast. Nevertheless, with the results of the study, there is still a need to improve the different intervention plans of the authority in alleviating the occurrence of the disease though it yielded a decreasing pattern in the future since it is evident that the figure of the forecasted data is still approximately 15,000 and above.

## 1. Introduction

Measles is a contagious viral infection directly responsible for yearly increased morbidity and mortality, particularly to children [1]–[3]. Common symptoms of measles in a person have a febrile illness with a rash, especially if they lack documentation of measles vaccination, a runny nose, a cough, red and watery eyes, and small white spots inside the cheeks [4].

Currently, the country is experiencing continuous outbreaks of measles [5]. Globally, there was an increasing trend of measles across all regions with the re-emergence of endemic and outbreaks in some countries that had previously achieved none occurrence level. World Health Organization defines measles elimination as the absence of endemic measles virus transmission in a defined geographical area.

Some countries are having high rates of measles vaccination experience outbreaks by virtue of imported cases causing transmission through susceptible groups of individuals who are not immune to the measles virus [6], [7]. In the United States, Netherlands, and United Kingdom, measles was declared eliminated where the disease is no longer constantly present in the country. Measles outbreak continues to occur in countries around the world, which was associated with importations from Israel. Reviews on national measles surveillance guidelines from Australia, Cambodia, Japan, New Zealand, Republic of Korea, and Canada were published on being free from endemic measles [3], [7], [8].

Several endemic countries have high incidence rates in Asia Pacific regions, including Laos, Malaysia, Vietnam, Indonesia, Thailand, and the Philippines [9], [10]. Measles cases in the Philippines are still the most numbered in the Asia Pacific. Based on the Department of Health (DOH) Measles and Rubella Surveillance report, there have been 43 214 measles cases reported [11]–[14].

## 2. Theoretical Framework

### 2.1. Review of Related Literature

Several literature and studies have started to explore data mining techniques in health and disease prediction [15].

Wang, Shen, Jiang, et al. applied ARIMA in predicting hemorrhagic fever with renal syndrome in China from 2011 until 2018 [16]. Mbau forecasted the number of lung diseases using ARIMA [17]. In the study conducted by He and Tao, the data from 2007 until 2015 was utilized to describe the influenza virus's epidemiology during the past nine influenza years [18]. Thus, the ARIMA model effectively predicts the positive rate of the influenza virus in a short time. Sato presented the process of using the ARIMA model on disease management, where the short period segments were encouraged to analyze each disease [19]. Several countries use the ARIMA model in forecasting the SARS-COV-2 diseases, including the Islamic Republic of Iran, Pakistan, Australia, [20]–[22]. Singh et al. studied the prediction of COVID-19 to 15 countries [23]. The model was used to forecast the estimated cases, deaths, and recoveries among the top fifteen (15) countries from April 24, 2020, until July 7, 2020. Some examples of the ARIMA model include predicting the beds that were occupied during the time where there was an epidemic of severe acute respiratory syndrome (SARS) at a hospital in Singapore [24]. Another study conducted in China recommended the need for an appropriate model to predict, using historical data like cases of hemorrhagic fever with kidney syndrome. Currently, China has 90% of cases of this disease reported globally, and the use of ARIMA models enables them to create better management and short-term predictions of the disease [20], [25], [26].

## 3. Materials and Methods

*Materials:* The data to be used in the study are situational report (SitRep) on measles outbreak taken from the Republic of the Philippines National Disaster Risk Reduction and Management Council (NDRRMC), Department of Health, Epidemiology Bureau, Public Health Surveillance Division, and International Federation of Red Cross and Red Crescent Societies. There were seventeen regions (17) in the Philippines, where each region has several cases and reported deaths. The historical data of reported confirmed cases of measles from 2015 until 2019 in the Philippines was utilized.

*Methods:* The current study utilized the Autoregressive Integrated Moving Average (ARIMA) model employed in many fields to construct models for forecasting time series [22], [27]. ARIMA(p,d,q) model is used to forecast the data pattern of diseases for the next fourteen years. Time-series predictions are basically based on the changes over time in historical data sets and can produce mathematical models by using statistical data that can be extrapolated [28]–[30]. The ARIMA (p,d,q) model is defined as follows:

$$X_t = \Phi_1 X_{t-1} + \ldots + \Phi_p X_{t-p} + a_t - \Theta_1 a_{t-1} - \ldots - \Theta_q a_{t-q} \quad (1)$$

Where, $\Phi$'s (phis) introduce the autoregressive parameters for estimation, $\Theta$'s (thetas) represents the moving average parameters, the main series is represented by X's, and the a's were the unknown random errors which are assumed to follow the normal probability distribution.

Three steps were performed to predict the measles cases by utilizing the ARIMA related modules. The model used auto-correlation analysis and partial auto-correlation analysis procedure to analyze random, stationery, and seasonal effects on time-series data. The proponent of the study prepared a stationary time series by considering the differences. Further, determined plausible models on the basis of an autocorrelogram and a partial autocorrelogram. Lastly, the parameter estimation and model testing were used to compare the plausible models obtained, and we selected the most appropriate model. Finally, we conducted a predictive analysis of historical data. The study used GRETL (Gnu Regression, Econometrics and Time-series Library) software for plotting the graphs and analysis of the data sets. Figure 1 presents the architectural design in predicting measles trends.

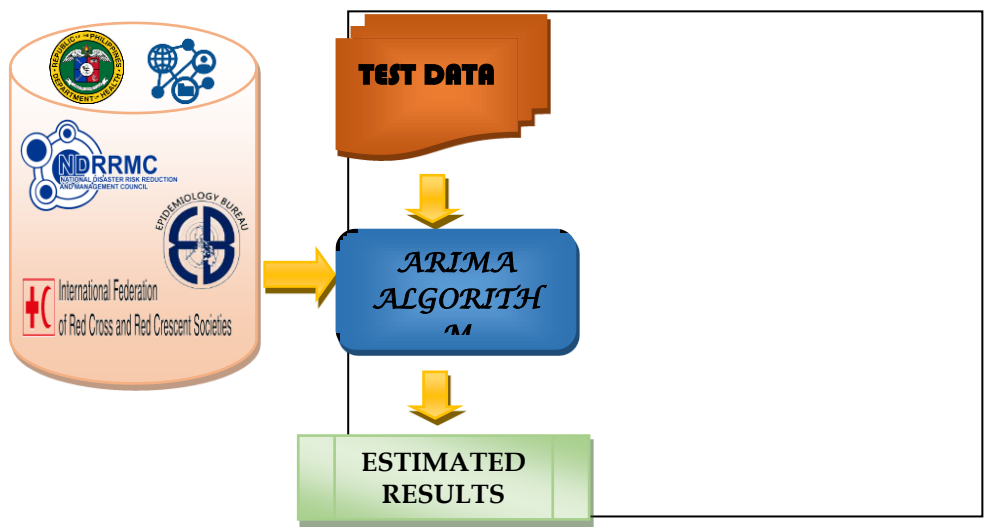

**Fig.1.** Predicting Measles Trends

## 4. Results and Discussion

The time-series plot showing the trend of reported confirmed cases of measles in the Philippines is presented in Figure 2. The 5-year historical data used was from the year 2015-2019. It is evident in the figure that there was a decreased pattern of the disease from the year 2015 to 2016.

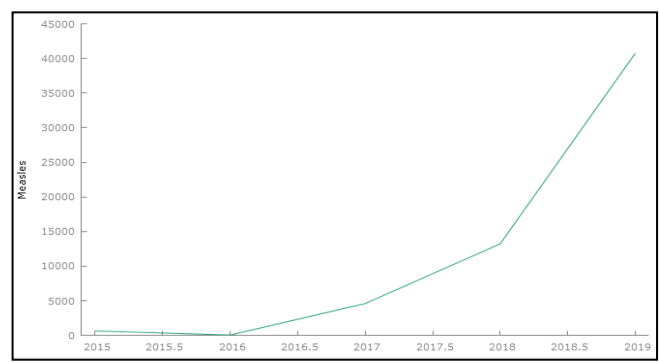

**Fig.2.** The Time-Series Plot of Measles in the Philippines

However, it is evident in the figure that there was a rapid increase in the disease based on the historical data from 2016 until 2019. The result is in line with the report of the WHO [6], [28] that there was an increase of cases, which is 30% from 2016-2017. Moreover, the result could be supported by the report of Mahase [10] that Measles cases have an evident increase by 300% so far in 2019 as compared with the same period in 2018 based on the data of the World Health Organization. Dyer [31] reported in his study that in the Philippines, 70 people, mostly children, died of measles in the first six weeks of 2019, based on the statement of the Philippines health ministry.

This came after 18 000 measles cases were reported by the country in 2018, a large increase from 2400 in 2017. Dyer [31] further noted that from 2014 to 2017, measles vaccination rates in the Philippines fell from 88% to 73% before dropping to 55% in 2018.

A five-year prediction data of Measles disease of the Philippines from 2020 to 2024 is presented in Figure 3 utilizing the 5-year historical data through ARIMA. It is evident in the figure that there will be a rapid increase of the disease in 2020, but eventually, it will be lessening from 2021 to 2024, showing a decreasing pattern of the disease based on the graph. Although, there is still a need for the health authority in the country to design effective mitigation plans focusing on the virus since the figure showed more than 15,000 forecasted cases in those years.

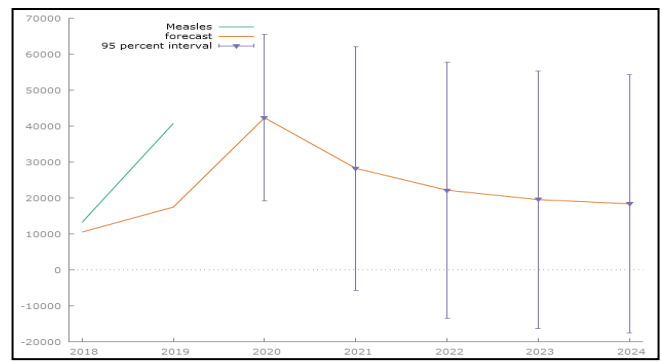

**Fig.3.** A Five-Year Forecast

## 5. Conclusions and Recommendations

Obtaining forecasted data on the occurrences of measles cases in very helpful to the administration in designing preventive measures against the spread of cases. In the study, the time-series tool successfully presented the trends of the disease in the country, and ARIMA (1,0,1) model helps in predicting the increasing and decreasing pattern of the measles cases. There was an increased pattern of the disease in the country from 2016 until 2019, based on historical data. However, in the next five years, there will be a decreasing number of its occurrence based on the results of the forecast. With the current situation and the future figure of measles disease in the country, there is still a need to strengthen the campaign by the Department of Health to the community in dealing with how to prevent oneself from the infection and at the same time to revisit previous intervention plans and programs and evaluate which part of the plans that need to be focused as a way to prevent that spread of the infections.